\documentclass[%
 reprint,
superscriptaddress,
 amsmath,amssymb,
 aps,
]{revtex4-2}

\usepackage{graphicx}
\usepackage{dcolumn}
\usepackage{bm}
\usepackage{hyperref}

\begin{document}

\preprint{APS/123-QED}

\title{Predicting multi-parametric dynamics of an externally forced oscillator using reservoir computing and minimal data}


\author{Manish Yadav}
\email{Corresponding author:\\manish.yadav@tu-berlin.de}
 \affiliation{Chair of Cyber-Physical Systems in Mechanical Engineering, Technische Universität Berlin, Straße des 17. Juni, 10623, Berlin, Germany}

\author{Swati Chauhan}
\affiliation{Department of Physics, Central University of Rajasthan, Ajmer, 305 817, Rajasthan, India}

\author{Manish Dev Shrimali}
\affiliation{Department of Physics, Central University of Rajasthan, Ajmer, 305 817, Rajasthan, India}
 %
\author{Merten Stender}
\affiliation{Chair of Cyber-Physical Systems in Mechanical Engineering, Technische Universität Berlin, Straße des 17. Juni, 10623, Berlin, Germany}%


\date{\today}

\begin{abstract}
Mechanical systems exhibit complex dynamical behavior from harmonic oscillations to chaotic motion. The dynamics undergo qualitative changes due to changes to internal system parameters like stiffness and changes to external forcing. Mapping out complete bifurcation diagrams numerically or experimentally is resource-consuming, or even infeasible. This study uses a data-driven approach to investigate how bifurcations can be learned from a few system response measurements. Particularly, the concept of reservoir computing (RC) is employed. As proof of concept, a minimal training dataset under the resource constraint problem of a Duffing oscillator with harmonic external forcing is provided as training data. Our results indicate that the RC not only learns to represent the system dynamics for the external forcing seen during training, but it also provides qualitatively accurate and robust system response predictions for completely unknown \textit{multi-}parameter regimes outside the training data. Particularly, while being trained solely on regular period-2 cycle dynamics, the proposed framework correctly predicts higher-order periodic and even chaotic dynamics for out-of-distribution forcing signals.\\
\begin{description}
\item[Key words: ]
critical transitions, bifurcations, minimal data, time series prediction, generalization.
\end{description}
\end{abstract}

\maketitle


\section{\label{sec1}Introduction}



Dynamical systems are omnipresent across the sciences, and are typically modelled mathematically by differential or difference equations for continuous or discrete time, respectively. Various scientific communities are interested in model-free methods capable of predicting the future state of dynamical systems from past observations, mostly motivated by systems for which no sufficiently accurate mathematical model exists yet. Well-known examples range from weather forecasting\cite{wu2024data} to stock market prediction \cite{wang2021stock} and epidemic spreading \cite{ghosh2021reservoir}, etc. Model-free forecasting models must be able to capture fast-scale instantaneous dynamics as well as slow-scale global dynamics (the so-called climate). Long-term time series predictions are challenging for methods that lack inherent correction mechanisms: classical auto-regressive methods employ a self-feeding feedback loop that advances in time by taking previous predictions as input for current predictions. Error accumulation is inevitable in those cases, limiting the prediction horizon, and potentially exhibiting stability issues.

Dynamical systems exhibit qualitative changes in their dynamics under variations of system parameters, typically denoted as bifurcations. Depending on the dynamical system at hand, parameter-induced transitions are critical, as they trigger unwanted or dangerous dynamics. A classical sequence of transitions can be found in the period-doubling route to chaos. Increasingly smaller parameter variations trigger transitions to higher-order periodic dynamics before the onset of chaotic motion. Predicting critical transitions in nonlinear dynamical systems is one of the pivotal challenges in the study of complex systems. Transitions, potentially in multi-dimensional parameter spaces, should therefore be considered and correctly predicted by forecasting methods that aim at relevant applications. Another level of complexity is introduced by multi-stability, which denotes the possibility of different co-existing stable system responses. Depending on the initial system state or instantaneous perturbations, other dynamics are observed for the corresponding basins of attraction. 

Recently, data-driven models from the wide field of machine learning have been applied to nonlinear dynamics forecasting tasks. While these models are generally very capable, the aforementioned critical transitions require comparable large data sets that resolve as many transitions as possible. The generalization capabilities of the model-free methods are then challenged not only by short- and long-term predictions for unseen system states, but also by unseen system parametrizations which may trigger unseen dynamics. A well-generalizing model should therefore be able to capture bifurcation structures in the training data and make acurrate predictions for unseen regimes of the bifurcation parameter space. This objective typically requires vast data sets with well-sampled parameter spaces, which is in contradiction with realistic forecasting scenarios. Here, only limited data is available, and no fine-grained sampling of bifurcations can be guaranteed. As a result, state-of-the-art deep learning architectures such as transformers, and generally extremely over-parametrized neural architectures, face severe challenges due to their data hunger \cite{oguntimilehin2014review}.   

As an alternative to deep neural architectures, reservoir computing (RC) has become a prominent method for predicting low-dimensional chaotic systems in recent years \cite{jaeger2001echo, borra2020effective, mandal2022machine, mandal2021achieving}. At its core, RC methods employ an unstructured dynamical graph to compute time-dependent input representations. During training, the trainable weights of a linear read-out layer are found by mapping the reservoir states to the given targets. The typical setup will learn a discrete flow map that advances the current system state (inputs) one time step into the future (outputs). Once the training is complete, the system transitions to a closed-loop configuration by connecting its output back to its input, enabling the method to evolve its state over time autonomously, without needing further input data. In this state, a well-trained reservoir computer can accurately predict the evolution of the original dynamical system, effectively functioning as a digital twin \cite{kong2023reservoir}. Recent studies have shown RC can accurately predict critical transitions, when trained with data from the pre-critical regime of the target system, thus effectively learning the dynamical climate of the target system \cite{panahi2024adaptable}. In another work, the authors employed reservoir computing to forecast various dynamics of a multistable system, where the learning method can capture the dynamics of the co-existing attractor after being trained with the single attractor’s dynamics \cite{roy2022model}.

In practical scenarios, the target system often encounters external driving forces, such as in the cases of driven lasers, regional climate systems \cite{barnett2005detecting}, aircraft engines, wind turbines, or ecosystems subject to environmental disturbances \cite{turner2003disturbance}. Motivated by the need mentioned above for minimal, energy-efficient, and model-free models with strong multi-parametric generalization properties, in this paper, we propose an auto-regressive reservoir computing framework to predict the response of an externally driven dynamical system across multiple unseen parameters and dynamics. Based on a minimal dataset, the proposed framework is tested to generate the full bifurcation diagram of the forced Duffing oscillator. A detailed study is conducted to identify the dependency of the number and parametric regimes of the drawn samples on the generalization properties of the auto-regressive RC. For that, multiple quantification measures are defined to evaluate the correspondence between the original and predicted bifurcation diagrams based on response amplitudes, periodicity, entropy and positing of multiple bifurcating points. Furthermore, we extend the study to predict a multi-parametric bifurcation diagram from minimal training samples. We identify that auto-regressive-RCs can predict multi-parametric bifurcation diagrams with very good precision of multiple transition points. Therefore, the multi-parametric generalization property of the reservoir computing framework will be a significant advancement toward minimal datasets and methods capable of generalizing to unseen dynamics. 

To accurately replicate the target system, the digital twin must include a mechanism that allows it to control or steer the neural network’s dynamics, thereby accounting for these external influences. This capability ensures that the reservoir computer can maintain an accurate representation of the target system's behaviour even under varying external conditions.

\section{Results}\label{sec2}

 The Duffing oscillator is considered for this study whose position ($q_1$) and velocity ($q_2$) are given by the following Eq. \ref{eq_DuffOsc}. It is described by a non-linear second-order differential equation that is commonly used to model certain damped and driven oscillators \cite{StrogatzBook, NLD_Thompson_Book}

\begin{gather}
\begin{aligned}
\Dot{q_1} =& q_2 \\
\Dot{q_2} =& -cq_2 - kq_1 -\beta q_{2}^{3} + f\cos(\omega t + \phi) \quad .
\end{aligned} 
\label{eq_DuffOsc}
\end{gather}

\begin{figure*}[htb]%
\centering
\includegraphics[width=0.92\textwidth]{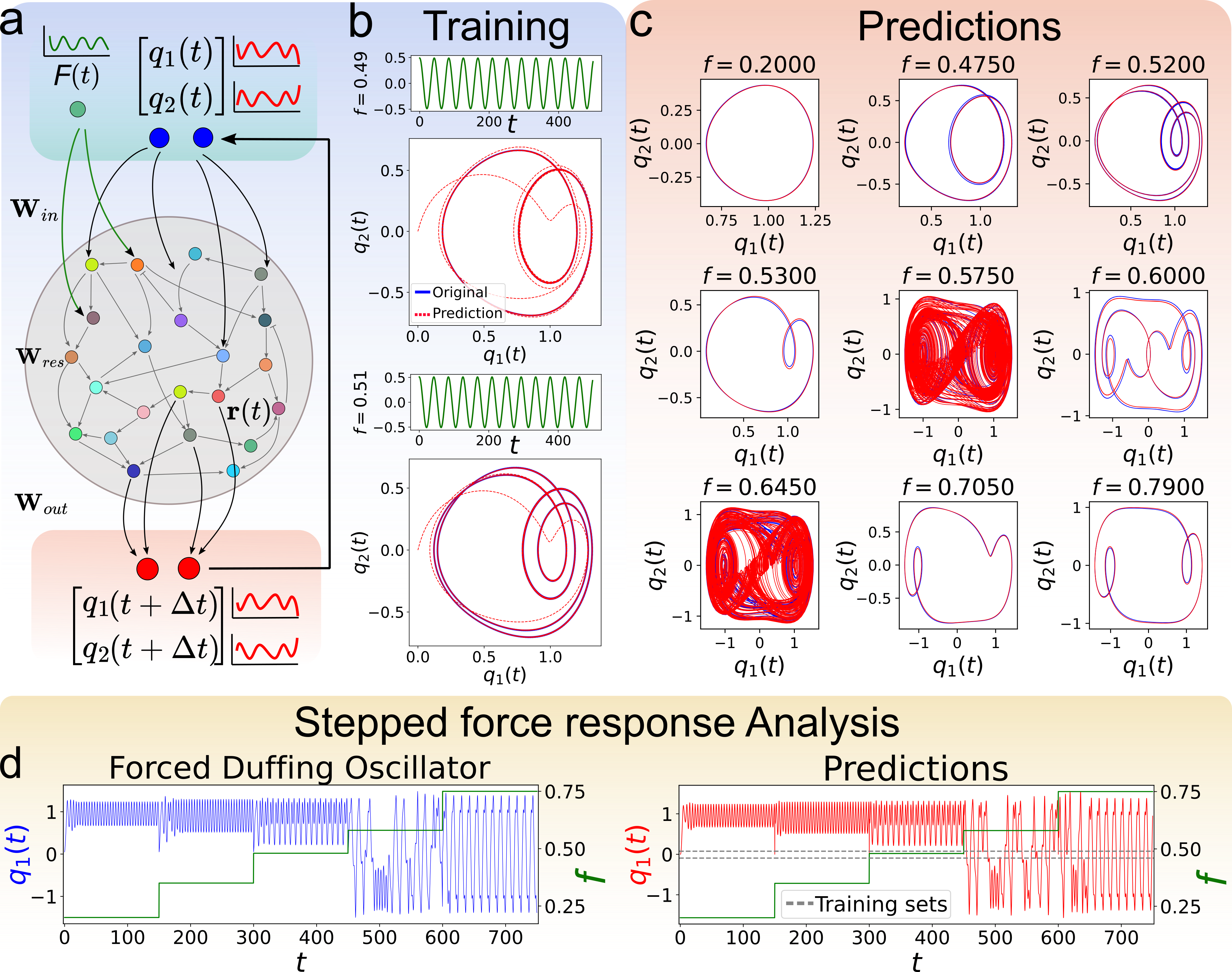}
\caption{\textbf{An auto-regressive reservoir computing method for predicting system dynamics for unknown parameters} (a) Schematic of the auto-regressive RC framework. (b) Training sets consist of two values for the forcing amplitudes $f=[0.49, 0.51]$. Shown are corresponding trajectories predicted by RC (\textit{red}) compared to the original trajectories (\textit{blue}). (c) Predictions of the dynamics for distinct forcing amplitudes ($f\in[0.1, 0.8]$) outside the training set. (d) (\textit{left}) The response of the forced Duffing oscillator to the stepped force and (\textit{right}) the response generated by the trained RC model. The amplitude of the stepped-force or $f$ is represented with \textit{green} lines. The external forcing amplitudes of the training set are represented with grey dashed lines.}\label{fig1}
\end{figure*}

\begin{figure*}[t]%
\centering
\includegraphics[width=0.9\textwidth]{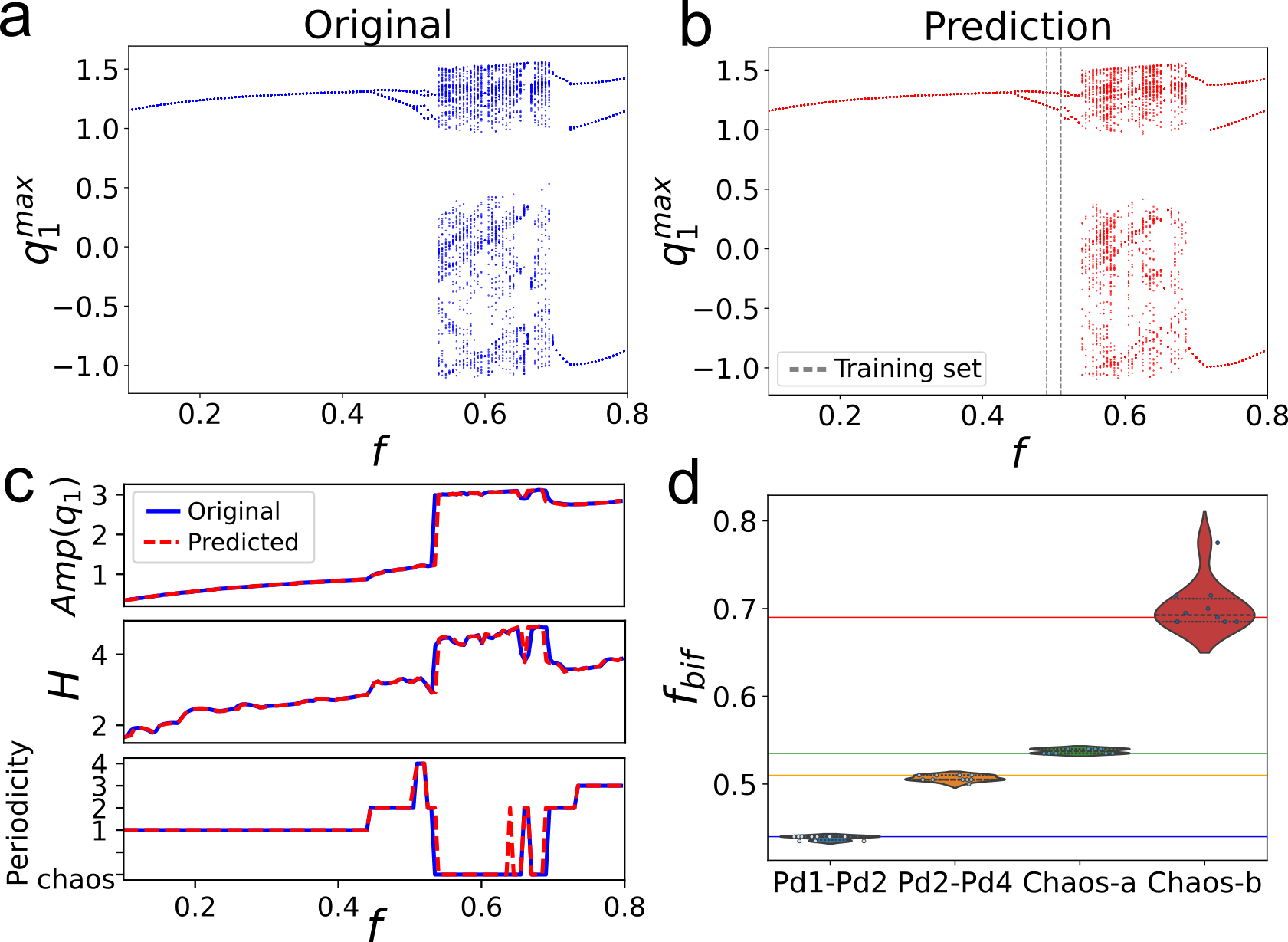}
\caption{\textbf{Bifurcation diagrams and their quantification.} Bifurcation diagrams derived from (a) the analytical oscillator, and from (b) the RC time-series predictions with the amplitude of the external forcing $f$ being the bifurcation parameter. Four different time series and bifurcation quantification measures are shown in (c) and (d): The oscillation amplitude (c, \textit{top}), entropy (c, \textit{middle}), periodicity (c, \textit{bottom}) and the transition points (d) (over 10 different RC models). The solid lines in panel (d) represent the original bifurcation points for the respective transitions.}\label{fig2}
\end{figure*}

The parameters \( c \), \( k \), \( \beta \) with respective values $[0.32, -1.0, 1.0,]$ represent the structural parameters of linear damping, linear stiffness, and non-linearity stiffness constant, respectively, when considering the Duffing oscillator in a mechanical context. The external harmonic force $F(t)=fcos(\omega t + \phi)$ has parameters frequency $\omega$ and phase-shift $\phi$, with respective values of $[1.5, 0.0]$. For this set of parameters, the Duffing oscillator showcases a rich dynamical response to varying forcing amplitudes $f$, resulting in single and multi-periodic cycles, and a period-doubling route to chaos; see Fig.~\ref{fig1}(c) and Fig.~\ref{fig2}(a). This system is well-suited for building a digital twin of a dynamical system with critical transitions that need to be learned from a minimal set of observations.

\subsection{\textbf{Auto-regressive reservoir computer for predicting appearance of chaotic dynamics}}

An auto-regressive version of the reservoir computer is employed in this study which is a combination of an autonomous and a non-autonomous RC (Fig.\ref{fig1}(a)), which means that it consists of an input that is externally driving the RC in a non-autonomous manner. Furthermore, since the RC framework is employed as a digital twin of the Duffing oscillator, the output layer is connected to the input layer for autonomous time stepping. In other words, the predicted [$q_1(t + \Delta t)$, $q_2(t + \Delta t)$] at the output layer are feedback to the input layer and the external periodic forcing $F(t)$ will non-autonomously drive the system. The reader is referred to the Methods Section~\ref{Method_auto-regressiveRC} for more details on the reservoir computing methods employed for this study. \\

To stick to the minimal dataset-based RC design, a training set consisting of only two samples is initially considered for generating the auto-regressive-RC-based digital twin of the forced Duffing oscillator. Hereafter, we refer to a \textit{sample} as a short time series of the system response to a given input with constant forcing values. Each sample consists of a total of 6500 timesteps. Two different forcing $f=[0.49, 0.51]$ amplitudes are selected that comprise the training set, generating period-2 and period-4 cycles, respectively. 

Fig.\ref{fig1}(b) displays the RC predictions (\textit{red}) for the training set samples in comparison to the original (\textit{blue}) trajectories. Starting from the very same initial conditions as the ground truth, the RC predicted trajectories follow the very same transients and settle to a period-2 and period-4 cycle, respectively. Neither qualitative nor significant quantitative differences between prediction and ground truth can be observed, corresponding to low training errors. For validation, the RC predicts the system response for unseen external forcing amplitudes $f$. Particularly, amplitude values close to the training set, but also far away from the training set are investigated. The validation cases cover period-1, 2, 3, and 4 cycles, and chaotic trajectories, hence dynamics that the model has never seen during training. The comparative study is shown in Fig.\ref{fig1}(c). It can be observed that the RC performs very well for out-of-sample regimes, i.e. for dynamics not covered in the training set, and forcing amplitude values far from the trained ones. The RC generalizes from the observation of a period-2 and period-4 data sample to other periodic cycles and chaos with surprising accuracy.

A parametric system response analysis is carried out to evaluate the instantaneous and steady-state predictions of the trained RC model for a step-like external forcing $f$ series. For that, the RC was driven continuously for a time of $T=150$ with a constant forcing of amplitude $f$, after which the amplitude instantaneously jumps to a higher value and remains there for the next $T=150$ steps, as shown with green in Fig.\ref{fig1}(d). The predictions of the auto-regressive-RC (\textit{red}) match the ground truth (\textit{blue}) in the steady-state regime, i.e. after the decay of some transients introduced by the jumps in external forcing amplitude. The transients differ between truth and predictions but decay quickly.

This initial study showcases the extrapolation and generalization capabilities of the proposed auto-regressive-RC framework that can predict the system behavior not only for dynamics similar to the training set but also qualitatively different dynamics. Moreover, the predictions qualitatively match the behavior of the Duffing oscillator in different dynamical regimes as the forcing amplitude is varied as a bifurcation parameter. This predictive performance was achieved by a training set comprising only two samples. Next, we will systematically study the dependency of the number of training samples on the generalization capabilities of the RC by quantitatively comparing its response and bifurcation diagrams to that of the analytical Duffing oscillator. Another point of interest is the selection of favorable dynamics to be selected for training that allows for maximum generalization. 

\subsection{\textbf{Learning 1D bifurcation diagrams}}

 A bifurcation analysis of the RC predictions is explored to qualitatively analyze the generalization for different external forcing $f$ marking the distinct dynamical regimes of the Duffing oscillator. The original bifurcation diagram of the forced Duffing oscillator is obtained by plotting the local maxima of the steady-state system response $q_1$ for changing $f$ values in range $[0.1, 0.8]$ as shown in Fig.\ref{fig2}(a). Five distinct dynamical regimes can be observed for different forcing amplitude regimes, see Table~\ref{tab:dyn_regimes}. Similarly, the bifurcation diagram obtained from the predicted trajectories by the auto-regressive-RC is shown in Fig.\ref{fig2}(b). Notably, the RC model is the same as in the previous study, being trained on only two samples. The maximum values are obtained from the steady-state response of the RC predictions for each amplitude value. The bifurcation diagram of the predictions qualitatively matches the one of the analytical model and captures all dynamic regimes.

\begin{table}[htbp]
    \centering
    \begin{tabular}{|c|c|c|}
    \hline 
         regime & forcing amplitude $f$& dynamics  \\
        \hline
         1 & $\left[0.1, 0.44\right]$ & period-$1$ cycle \\
         2 & $\left(0.44, 0.51\right]$ & period-$2$ cycle \\
         3 & $\left(0.51, 0.535\right)$ & period-$4$ cycle \\
         4 & $\left[0.535, 0.69\right]$ & chaos \\
         5 & $\left(0.69, 0.8\right]$ & period-$3$ cycle \\
    \hline
    \end{tabular}
    \caption{Dynamical regimes of the studied forced Duffing oscillator}
    \label{tab:dyn_regimes}
\end{table}

         

The predicted dynamics are quantitatively compared with the original ones using four different measures covering the amplitude of motion, its periodicity, the entropy, and the location of bifurcation points. See Methods Sec. \ref{Method_Quantify} for details of these quantification measures. The amplitude of $q_1(t)$, entropy $H$, and the periodicity values of the predicted trajectories (\textit{red}) are compared directly with those of the original ones (\textit{blue}) in Fig.\ref{fig2}(c). Along the bifurcation parameter $f$ there are only very few points at which the measures deviate: not only the amplitude of oscillation, but also its period and the entropy match almost perfectly in all dynamic regimes. Slight deviations in the chaotic regime around $f=0.7$ can be observed. It becomes evident that the predicted trajectories and the bifurcation diagrams not only qualitatively but also quantitatively match the original ones generated by original Duffing oscillator. Trained on two samples of period-2 and period-4 dynamics, the RC can predict chaotic motion for other forcing values with correct entropy, and other periodic dynamics with correct amplitude and periodicity.  

Furthermore, the appearance of bifurcation points between different dynamical regimes, namely from cycles of period 1 to 2 (Pd1-Pd2), from period 2 to 4 (Pd2-Pd4), the beginning of chaos (Chaos-a) and the end of the chaotic regime (Chaos-b) are tracked. To introduce some robustness, these four different bifurcating points ($f_{bif}$) are obtained from 10 different RC models (trained independently) and they are shown in Fig.\ref{fig2}(d) where solid lines reflect the original values of the corresponding bifurcation points.\\

 Now, the apparent questions that arise from these observations are: \textit{what is the minimal data limit within which the proposed auto-regressive-RC model can accurately predict the system trajectories and the consequent bifurcation diagrams? Which dynamical regimes are particularly suited for training?}




\subsection{\textbf{Data set requirements}}

\begin{figure*}[htb]%
\centering
\includegraphics[width=0.92\textwidth]{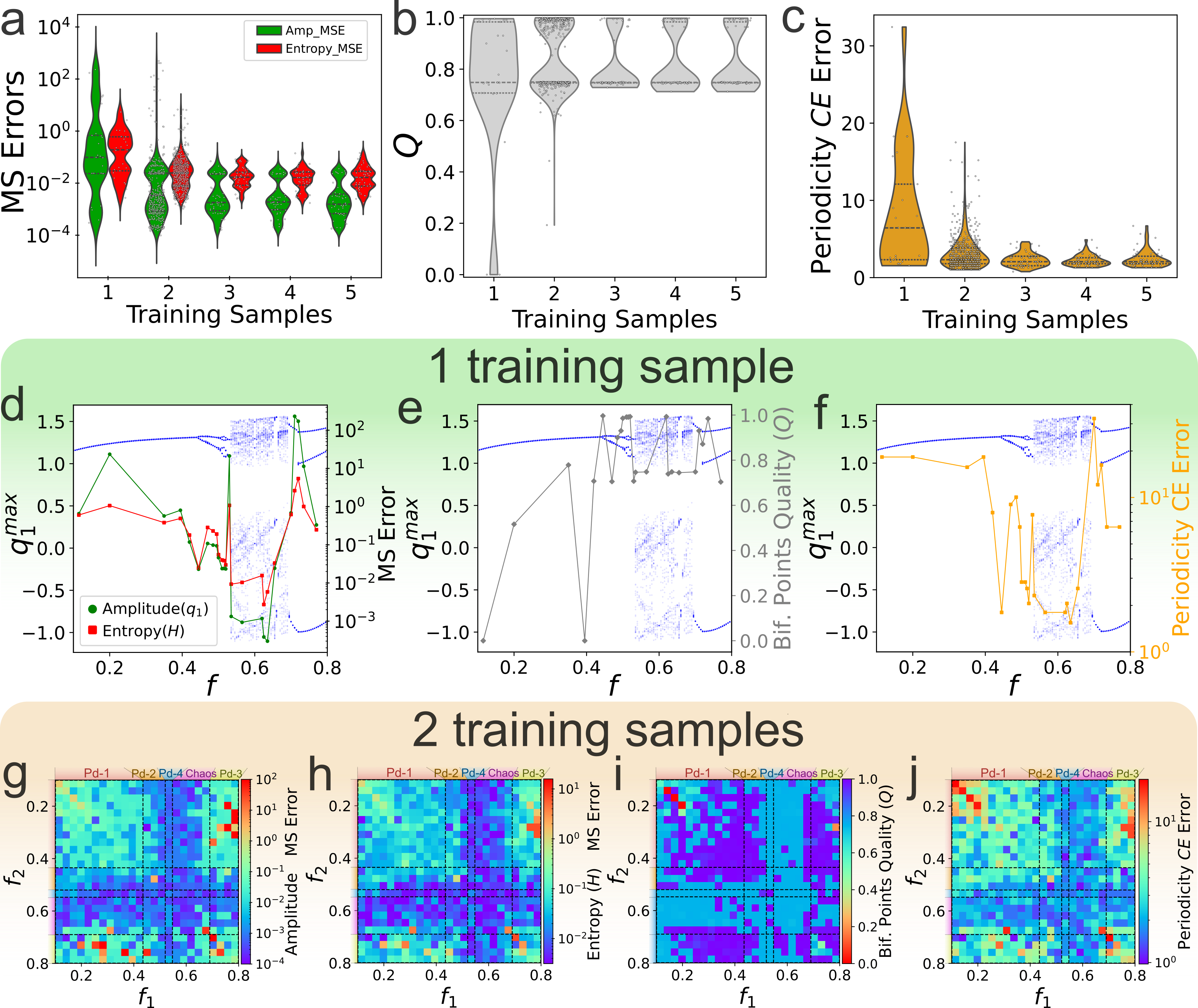}
\caption{\textbf{Bifurcation diagram prediction with various samples.} 
auto-regressive-RCs were trained with samples $\in[1, 5]$ and the obtained bifurcation diagram over $f$ are quantified relative to the original one.
The Mean Square (MS) errors of Amplitude and Entropy($H$), the Bifurcation Points Quality ($Q$) measure (b) and the overall Cross Entropy (\textit{CE}) error of the periodicity over $f$ are respectively shown in panels (a), (b) and (c). The detailed quantifications of the predicted bifurcation diagrams obtained from auto-regressive-RCs trained with only a single sample are shown in panels (d)-(f). All 4 quantifiers are plotted against the $f$-value from where that sample was drawn. The original bifurcation diagram (\textit{blue}) is for the reference. (g)-(j) The quantification of the predicted bifurcation diagrams obtained from auto-regressive-RCs trained with 2 samples $[f_{1}, f_{2}]$. \textbf{Note:} The deviations of each bifurcation point are shown in Supplementary Fig.\ref{SupFig3_d} that are then used to calculate the $Q$ measures shown in (e) and (i) using Eq.\ref{Eq_BifQuality}.
}\label{fig3}
\end{figure*}

In the previous section, a minimal data set composed of two training samples suffices to learn a complex transition behavior and extrapolate to unseen dynamics including chaos. One research question that arises from those results is: \textit{At which parameters should the training data be sampled, and do more training samples increase model performance?} The training data for the results in Fig.\ref{fig1}(b) was carefully chosen to represent a period-$2$ cycle at $f_1=0.49$ and a period-$4$ cycle at $f_2=0.51$, thus capturing different dynamics in the training set.

Experiments are conducted to examine the impact of larger and more diverse training datasets by progressively increasing the number of samples. Beginning with a training set of a single sample, a total of $5$ samples are generated by randomly selecting (without replacement) from the full bifurcation parameter range $f\in \left[0.1, 0.8\right]$, and the corresponding prediction errors are recorded. For training samples of more than one at least $50$ independent experiments are performed. \\

\textbf{One-sample training sets} were generated by randomly drawing five external forcing values $f$ from each dynamic regime shown in Table~\ref{tab:dyn_regimes}, making a total of 25 training sets. The 25 independent RC models trained on different single samples showcase a large diversity in the quantification measures of predicted trajectories and bifurcation diagrams with high error values of amplitudes, entropies ($H$) and periodicity cross-entropy (CE) (Fig.~\ref{fig3}~(a) and (c)) and the bifurcation points quality measure $Q$ (Fig.\ref{fig3}(b)). In order to understand which training sample in terms of dynamics regime provides worst and best predictions, these error measures (mean square error across the complete $f$ range validation set) are plotted against the $f$ values from which that single training sample was drawn (Fig.\ref{fig3}(d)-(f)). The mean square errors of the trajectory amplitudes, entropy (Fig.\ref{fig3}(d)), and cross-entropy error of periodicities (Fig.\ref{fig3}(f)) are remarkably low in the chaotic regime as compared to the other periodic regimes. This means that the models trained with samples drawn from chaotic regimes performed best while those from periodic regimes showcased high errors or low values of bifurcation points quality $Q$. \\

\textbf{Two-sample training sets} were generated using all possible combinations of $f$ values used in the single-sample case, creating a total of 625 training sets. 625 corresponding RC models are trained, and evaluated for the complete range of forcing amplitudes. Each of the four quantification measures for the entire bifurcation diagram is shown in the 2D parametric space of $[f_1, f_2]$ (Fig.~\ref{fig3}\,(g)-(j)) from where that sample was drawn. In other words, the coordinate in $[f_1, f_2]$ represents the training sample for that RC model and its performance is represented by color. The distinct dynamic regimes are marked with dashed lines.

It is evident that those models trained with at least one sample drawn from the chaotic or Pd-4 regimes perform better (\textit{deep purple} color) than those containing none of these dynamics. However, it is worth noting that whenever both the $f$ values belong to chaotic regimes, the Pd-4 cycles are completely missing from the predictions (see Supp. Fig.\ref{SupFig3_d}(f)). This is not the case when one of the two samples is drawn from Pd-1 and another from Pd-2. This explains why the resulting $Q$-measure values are inverted in Fig.\ref{fig3}(i) when compared to other quantifiers.\\

Overall, better quantification measure values are observed for the predictions when the training set size is increased from one to two, which is also visible in the first row of Fig.\ref{fig3}.\\

\textbf{Multiple-sample training sets} were furthermore generated to elucidate the dependency of the prediction quality on the number of training samples. 50 different training sets were drawn of sample sizes 3, 4 and 5 (each without replacement). The four different quantification measures of the predicted trajectories and the entire bifurcation diagrams are represented in Fig.\ref{fig3} (a)-(c) for training sets consisting of 1 to 5 samples each.

The quality of predicted trajectories and the entire bifurcation diagrams improves by increasing the training samples from 1 to 2, which is quantified by lower mean square values of amplitude, entropy and periodicity errors (Fig.\ref{fig3} (a) and (c)) or high bifurcation points quality $Q$ values (Fig.\ref{fig3} (b)).
Furthermore, it is also evident that by increasing the training sample size by more than 2, the overall quality of the predicted trajectories and the bifurcation diagrams does not improve much as compared to that observed in training samples of size 2.\\

This study therefore exhibits the generalization capabilities of the auto-regressive RC models under very limited training data. The predicted bifurcation diagram with at least two training samples over a single parameter ($f$) turns out to be quantitatively matching the original one. Moreover, after evaluating different performance quantifiers for 1 and 2 sample-trained auto-regressive-RCs, the Pd-2 regime emerges as the optimal parametric window for accurately predicting system responses in unknown parameter regimes. This observation leads to a consequent question: \textit{is a multi-parametric generalization of a dynamical system also possible in a auto-regressive-RC framework?}

\subsection{\textbf{Learning 2D bifurcations: variation of forcing amplitude and frequency}}

\begin{figure*}
\centering
\includegraphics[width=0.88\textwidth]{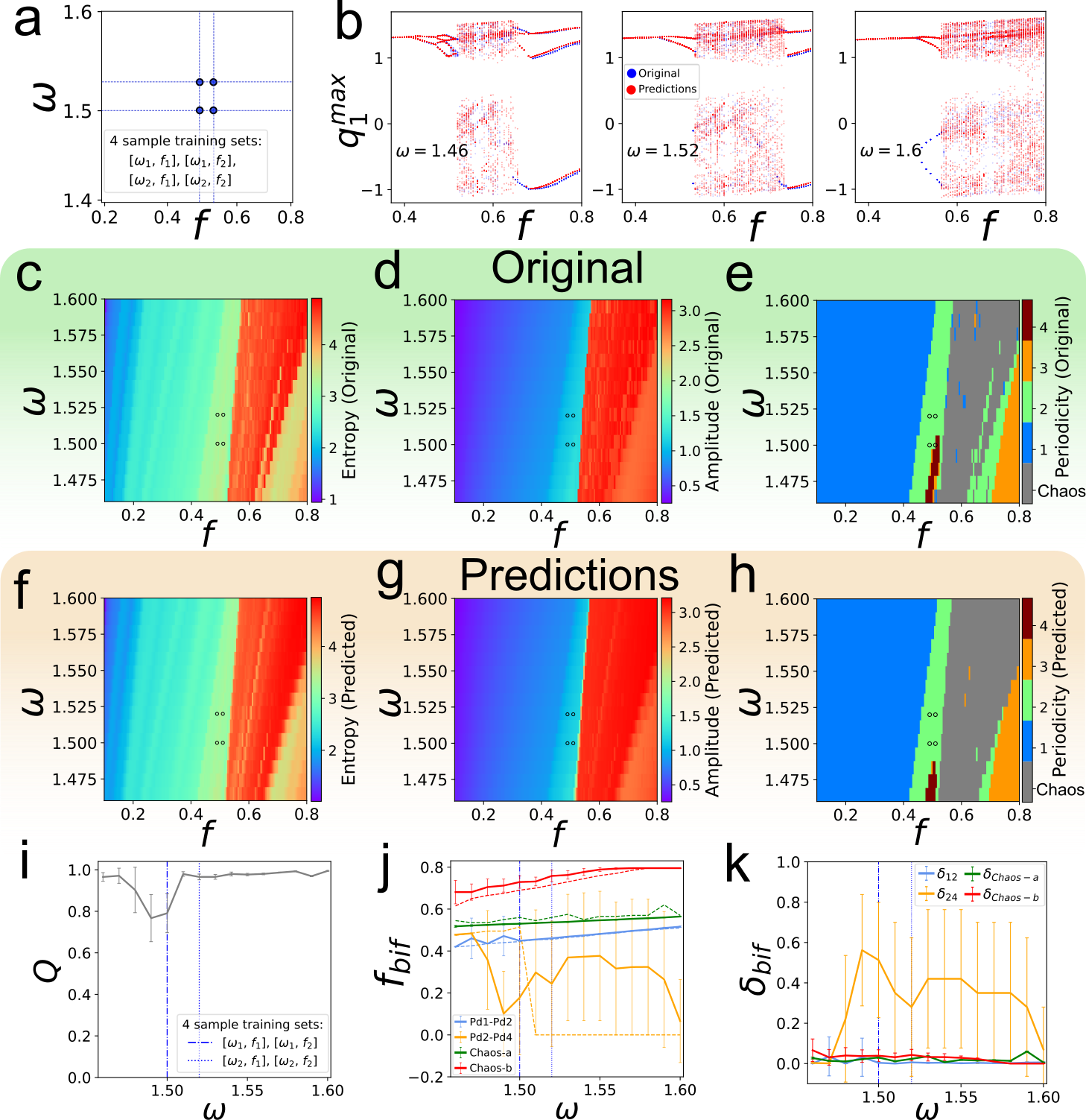}
\caption{\textbf{2D Bifurcation diagram prediction with 4 sample training dataset.} (a) Schematic showing the selection of a 4-sample training set in the 2-dimensional parametric space of $[f, \omega]$. (b) The changing landscape of the bifurcation diagram for increasing $\omega$ values from panels \textit{left} to \textit{right} in b) and the corresponding predictions by the RC model. The entropy, amplitude, and periodicity of the original ((c)-(e)) and the predicted ((f)-(h)) dynamics are shown for the full 2D bifurcation space. (i) displays the bifurcation points quality ($Q$) measure over external forcing $f$, (j) the transition points (original are shown with dashed lines), and (k) their deviations ($\delta_{bif}$) relative to the original transition points are shown concerning the changing values of $\omega$, and the variations stemming from 10 different RC models.}\label{fig4}
\end{figure*}

The forced Duffing oscillator shows rich dynamic behaviors for varying forcing frequencies. The entire bifurcation diagram shifts towards higher amplitudes $f$ with increasing forcing frequency $\omega$ (Fig.\ref{fig4}(b) \textit{blue}). The oscillator shows windows of periods 1, 2 and 4 followed by chaos and then period 3 cycles appear for low $\omega$ values. However, on increasing $\omega$ some high periodic windows disappear for instance, period-4 cycles do not arise for $\omega = 1.52$. Further, on increasing $\omega$ even more, period-3 cycles also disappear and chaotic trajectories cover the majority of the $f$ parametric space. The detailed quantification of the 2D bifurcation diagram of the forced Duffing oscillator reveals the changing dynamical regimes in the $[f,\omega]$ parameter space (Fig.\ref{fig4}(c)-(e)). This makes the forced Duffing oscillator an ideal system to test the multi-parametric generalization capabilities of the auto-regressive-RC framework in the 2D parametric space of $[f, \omega]$. 

To equip the readout layer of the RC with generalization capabilities over signal frequency ($\omega$) along with its amplitude $f$, a four-sample training set is generated in the $[f,\omega]$ space at $f=[0.49, 0.51]$ and $\omega=[1.50, 1.52]$ (schematically shown in 
 Fig.~\ref{fig4}(a)). The bifurcation diagrams derived from the predictions are compared against the ground truth for three validation frequency values in Fig.~\ref{fig4}(b). Not only is the RC model capable of predicting the correct qualitative dynamics along the forcing amplitude variation, but it also nicely captures the global shift of the bifurcation structures under change in the excitation frequency value \cite{2DBifVideo}. Thus, the RC generalizes from 4 training samples to a very accurate resolution of complicated bifurcation structures in a two-dimensional parameter space, including regular and irregular dynamics and several bifurcations. 
 
 Fig.~\ref{fig4}(c) to (h) display the quantitative analysis of the predictions in terms of entropy, motion amplitude, and periodicity. The complete two-dimensional parameter space predictions match the ground truth characteristics very well. Particularly the regular dynamics in lower $f$ regimes are captured very well. Slight deviations can be observed for high $f$ values, where the dynamics change from regular to irregular under the variation of $\omega$. Tongues of different periodicities are captured well by the RC model trained on only four training samples. Moreover, the transition between different dynamical regimes, for instance, the transition from period-1 to period-2 marked by the sudden change of entropy (\textit{yellow} to \textit{red} in Fig.\ref{fig4}(f)), amplitude (\textit{cyan} to \textit{red} in Fig.\ref{fig4}(g)) and periodicity (\textit{sky-blue} to \textit{green} in Fig.\ref{fig4}(h)), was precisely predicted by the RC models. The prediction of the dynamical regimes is even more interesting from the perspective of predicting critical transitions.\\

Quantitatively, the accurate transition points are captured by the quality metric $Q$, for which higher values indicate better predictions. To account for the variation introduced by the random generation of reservoir networks, 10 different RC models were trained independently. Fig.\ref{fig4}(i) displays the quality parameter $Q$ along the forcing frequency. For frequency values greater than $\sim1.525$, the predictions are very accurate. A drop can be observed for smaller values i.e $\sim1.5$, especially corresponding to prediction errors for the transitions from period-2 to period-4 and then eventually to the chaotic regime. The variations stemming from different reservoir networks of the same size are notable for the less accurate predictions, and vanishing for the accurate predictions.

Only the transition point from period-2 to period-4 was not precisely predicted by the auto-regressive-RC models, thereby generating large deviation in $\delta){24}$ after accurate predictions in initial $\omega$ values (\textit{yellow} curves in Fig.\ref{fig4}(j) and (k)). This resulted in a drop in the $Q$-measure for those $\omega$ values (Fig.\ref{fig4}(i)). It is important to highlight that period-4 cycles are observed exclusively within the small region of $[f,\omega]$ space indicated by the \textit{brown} color in Fig.\ref{fig4}(e). Notably, the auto-regressive-RC models also predict this behavior to a significant extent, despite being trained on only 4 samples.

The 2-dimensional bifurcation diagram predicted by the RC models trained on a minimal training set consisting of only 4 samples are extremely accurate with the precise periods, amplitude and as well as entropy of the trajectories. Furthermore, the transition points were also captured by the RC models with great accuracy given the fact that the bifurcation diagram changes differently with both $f$ and $\omega$. Therefore, the autoregressive-RC design showcases the generalization property over the amplitude and frequency of the external forcing of the duffing oscillator.

\section{Discussion and Outlook}\label{sec3}


In this paper, we introduced an auto-regressive design of a reservoir computer to predict the behavior of an externally driven Duffing oscillator for unseen parameters, focusing on qualitative changes in dynamics. Using minimal data, we generated the complete bifurcation diagram of the forced Duffing oscillator. We analyzed how data quantity and sampling regimes impact the generalization of the auto-regressive RC. Various measures, including response amplitudes, periodicity, and entropy, were used to compare predicted and original bifurcation diagrams. Additionally, we extended the study to multi-parametric bifurcation prediction, demonstrating the RC’s accuracy and generalization with minimal training, advancing the development of efficient digital twins for complex systems. This demonstrates the multiparametric generalization potential of the RC framework, marking a significant step in developing green digital twins of complex dynamical systems with minimal datasets and training.\\

This research focuses on scenarios where only limited datasets are available, a common situation in experimental data collection. We also analyzed the performance of RC when both the frequency and amplitude of the driving force varied. Motivated by this, we implemented a ``parameter-aware" RC to capture the dynamics of a Duffing oscillator subjected to external periodic forcing. Our results demonstrate that a well-trained RC can effectively model the system's dynamics with minimal data when the system experiences changes in both frequency and amplitude.\\

One of the objectives of this research was also to develop a green digital twin that is data-efficient, easy to train, and capable of modeling complex systems. The proposed auto-regressive-RC framework meets these criteria, making it suitable for use as a digital twin for more complex systems, such as multi-body systems, in both research and industrial contexts. Additionally, the auto-regressive-RC design offers potential energy efficiency improvements by optimizing the \textit{reservoir} network. Techniques like the graph-theoretical measures and goal-oriented network evolution methods proposed in \cite{EvolveRC_Yadav} could be used to create specialized, smaller, and more efficient reservoir computers with strong generalization capabilities.\\

Our study contributes to a deeper understanding of RC's potential in dynamical systems, but there are opportunities for further research. For instance, applying this approach to dynamical systems exhibiting multistability under time-varying external forces should be explored. While our study focuses on a single RC due to the small input dimension, future work could involve implementing parallel RCs for large, complex systems \cite{pathak2018model}. However, certain limitations should be noted. Our study assumes access to all state variables, which may not be realistic for scenarios involving experiments. Additionally, real data often contains noise. Nevertheless, existing literature suggests that RC performance can be enhanced by adding an appropriate amount of noise to the data \cite{kong2023reservoir}.



\begin{acknowledgments}
This work was supported by the Deutsche Forschungsgemeinschaft (DFG, German Research Foundation) under the Special Priority Program (SPP 2353: Daring More Intelligence – Design Assistants in Mechanics and Dynamics, project number 501847579). MDS acknowledges financial support from SERB, Department of Science and Technology (DST), India (Grant No. CRG/2021/003301).
\end{acknowledgments}

\section{Methods}

\subsection{\textbf{Auto-regressive RC Design}}
\label{Method_auto-regressiveRC}
In this article, we implemented the Echo State Network (ESN) framework that was originally proposed by Jaeger \cite{jaeger2001echo}.  Typically, this type of architecture consists of three layers: an input layer, a reservoir and an output layer. We consider an m-dimensional time series data, accompanied by the time series of driven force. Thus input data consists of m+1 dimensional data represented as $[u(t) = \hat u(t)\quad F(t)]$, where $F(t) = f\cos(\omega t + \phi)$ with $f$ being the amplitude and $\omega$ the frequency of the driving force $F(t)$. The input signal is encoded into the reservoir with $N_{res}$  nodes using a $[(m+1) \times N_{res}]$-dimensional weighted matrix $\textbf{W}_{in}$. The weights of $\textbf{W}_{in}$ are randomly sampled from range $[-\sigma,\sigma]$, where $\sigma$ is one of the hyperparameters. The matrix $\textbf{W}_{in}$ is structured in such a way, that the information of the $m$-dimensional time series is uniformly distributed across all the nodes of the reservoir, which means each input dimension information is encoded into $\frac{N_{res}}{m}$ of nodes. However, the information of the driving force is encoded into all nodes of the reservoir, enabling the reservoir to learn the relationship between changes in system dynamics as the external driving force varies. The encoded information within the reservoir is projected into high dimensional space using the matrix $\textbf{W}_{res}$ an $(N_{res} \times N_{res})$ matrix. This matrix is also a random sparse matrix generated with the spectral radius $\rho$. The reservoir dynamics can be described by the following map:

\begin{equation}\label{eq:st_upd}
    \textbf{r}(t+1) = (1-\alpha)\textbf{r}(t) + \alpha~\mathrm{tanh}[\textbf{{W}}_{res}\cdot \textbf{r}(t) + \textbf{{W}}_{in} \cdot u(t)],
\end{equation}

where $\textbf{r}(t)$ represents the reservoir state at time t. Here $\alpha $ is the leakage parameter that takes the value from 0-1. It controls the trade-off between the non-linearity of input data and information of reservoir dynamics.
In the first case, we train the machine with the time series of the external driving force for different values of amplitudes ($f$) while keeping the frequency ($\omega$) fixed throughout training.  
In the second scenario, the neural machine is trained with the $N_{p}$ number of time series of the driving force, varying both the $f$ and $\omega$, thereby making the reservoir aware of both parameters simultaneously. For both the cases, the reservoir is trained over $N_{t}$ time steps and the reservoir states are stored after discarding some $T_{trans}$ transient steps. Then the remaining reservoir states are stored in a $\textbf{R}_{res}$ matrix with dimension [$N_{res} \times N_{p}(N_{t}-T_{trans})]$. Once the $\textbf{R}_{res}$ matrix is computed, we can generate the readout weight matrix using the linear equation:

\begin{equation}\label{eq:output}
    \textbf{U}= \textbf{W}_{out} \textbf{R}_{res}
\end{equation}

where $\textbf{U}$ is the matrix containing the target data sets, with dimension $[m  \times  N_{p}(N_{t}-T_{trans})]$. To minimize the difference between target output and $\textbf{W}_{out}\textbf{R}_{res}$, we apply Ridge regression and determine the readout layer using the following expression:

\begin{equation}\label{eq:readout_layer}
    \textbf{W}_{out}= \textbf{U} \textbf{R}_{res}^T[\textbf{R}_{res}^T\textbf{R}_{res} + \beta I ]^{-1}
\end{equation}

where $\beta$ is another hyperparameter that prevents overfitting. After completing the training phase, the neural machine is ready to make predictions. During the prediction phase, we introduce a new driving force, unseen during training and the machine produces the corresponding time series in an autonomous way using the following equations:

\begin{equation}\label{eq:pred_end}
    \textbf{r}(t+1) = (1-\alpha)\textbf{r}(t) + \alpha~\mathrm{tanh}[\textbf{W}_{res}\cdot \textbf{r}(t) + \textbf{W}_{in} \cdot \mathbf{v}(t)]
\end{equation}

\begin{equation}
    \mathbf{v}(t) = [  \mathbf{\hat v}(t) \quad Fcos(\omega t)]^{T}
\end{equation}

\begin{equation}\label{eq:predicted_output}
   \mathbf{\hat v}(t) = \textbf{W}_{out} \textbf{r}(t),
\end{equation}

By repeating the equations \ref{eq:pred_end}-\ref{eq:predicted_output} we store the predicted output after discarding a few reservoir's warmup iterations.  The above-mentioned scheme also involves some hyperparameters, namely the leaking parameter ($\alpha$), spectral radius ($\rho$), regularization parameter ($\beta$), and input scaling factor ($\sigma$). These hyperparameters need to be optimized before making the predictions which can be achieved using a simultaneous optimization technique \cite{griffith2019forecasting}-\cite{yperman2016bayesian} for the loss function defined as the root-mean-square error (RMSE) of the prediction. After obtaining an optimal set of hyperparameters, the machine is ready to predict the dynamics of the Duffing oscillator under different external forcing conditions.


\begin{figure*}[htb]%
\centering
\includegraphics[width=1\textwidth]{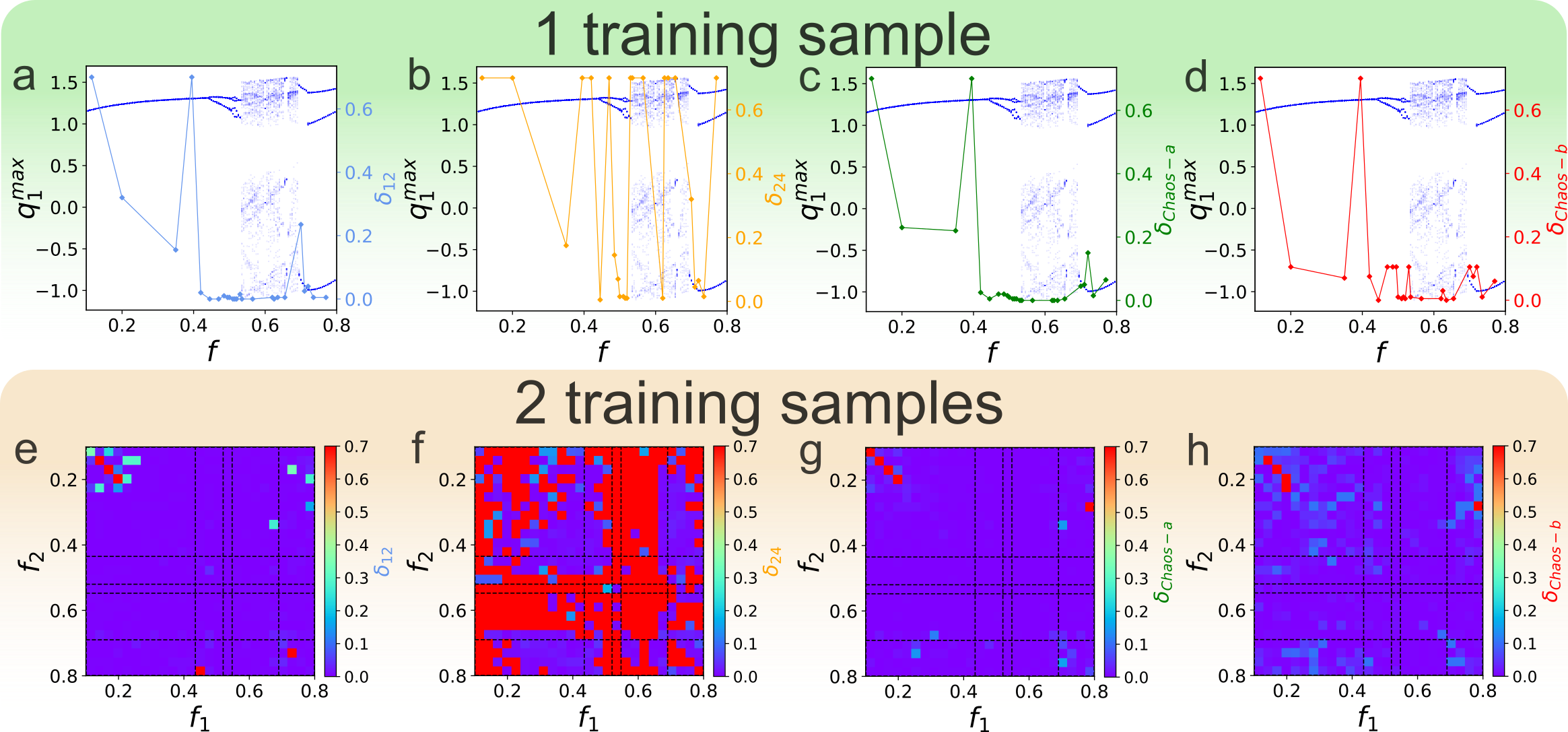}
\caption{\textbf{Error in bifurcation points predicted from 1 and 2 samples over $f$.} The bifurcation diagram in the background (blue) represents the dynamic regime from where the single sample is drawn. The bifurcating points Pd1-Pd2, Pd2-Pd4, Chaos-start and Chaos-end are tracked and their respective deviations ($\delta_{12}$, $\delta_{24}$, $\delta_{Chaos-a}$ and $\delta_{Chaos-b}$) from the original points are shown in respective panels from ($a$) to ($d$).}\label{SupFig3_d}
\end{figure*}

\subsection{\textbf{Quantification measures}}
\label{Method_Quantify}
The predictive performance of the auto-regressive-RC models is quantified using four distinct following measures:\\  

\textbf{(a) Amplitude:}
As a natural choice for measuring oscillatory data, the absolute amplitude $A$ of motion is obtained by calculating the difference between the global maximum and minimum of $ q_1$ variables.

\begin{equation}
    A=\left| \mathrm{max}\left(q_1 \left( t \right)\right) - \mathrm{min}\left( q_1\left(t\right) \right) \right|    
\end{equation}

this absolute amplitude $A$ of the system variable has been utilized for furthermore model evaluation using the mean square error between the original ($A$) and the predicted amplitudes ($\hat{A}$):

\begin{equation}
E_{\mathrm{A}} = \frac{1}{m} \sum_{i=1}^{m}\left(A_i - \hat{A}_i \right)^2, \quad 
\end{equation}

where the mean is over the number of model repetitions ($m$).\\ 

\textbf{(b) Entropy:} It is possible to verify the degree of phase space occupancy for the predicted dynamics by examining the entropy profile.  We utilize the Shannon entropy \cite{shannon1948mathematical} approach to analyze the entropy profile of the predicted and original time series with regard to the external force as depicted in Fig.\ref{fig2}(c). The entropy of a time series is computed via 

\begin{equation}
    H =\sum_{i=1}^{N}p_{i}log(p_{i}) 
\end{equation}

The computation is performed by discretizing the complete state space volume using appropriately tiny unit volumes, where pi is the nonzero likelihood that the ith phase space state will be filled. The total number of states with nonzero probability is N. We note that the entropy profile of the predicted dynamics closely resembles the original system's.

The entropy error can be calculated as:
\begin{equation}
    E_{\mathrm{H}} = \frac{1}{N} \sum_{i=1}^{N}\left(H_i - \hat{H}_i \right)^2  \quad .   
\end{equation}
This nonlinear invariant measure is especially useful for chaotic dynamics for which neither an point-wise Euclidean measure, nor a periodicity information is meaningful. \\

\textbf{(c) Periodicity}: To classify the predicted dynamics qualitatively, a Poincaré section is introduced at vanishing velocities $q_2(t)=0$. The intersection points of a given trajectory are clustered using unsupervised machine learning (DBSCAN, $5$ min. points, $\epsilon$ neighborhood chosen according to maximum silhouette coefficient via grid search) to obtain the periodicity $\gamma$ of the trajectory. If the trajectory intersects the Poincaré section twice, a period-$1$ cycle is detected. The periodicity can thus be used as a class label to compare true and predicted dynamics, and to detect transitions as a function of external parameters such as the forcing amplitude $f$. As the dynamical system at hand exhibits different types of regular dynamics up to period-$4$ cyclic dynamics, the evaluation is rendered a multi-class classification problem. Thus, the periodicity error is computed as categorical crossentropy 
\begin{equation}
    E_{\mathrm{CE}} = -\frac{1}{N}\sum_{i=1}^{N} \sum_{k=1}^{K} y_{i,k} \mathrm{log}\left(\hat{y}_{i,k}\right)
\end{equation} 
from the true $y_{i,k}$ and predicted $\hat{y}_{i,k}$ probability of the $i$\,th signal's periodicity, where $K$ gives the number of different globally observed periodicities. The periodicity error is only evaluated in regimes of regular dynamics, and not in regimes of chaos.\\

\textbf{(d) Bifurcation Points Quality ($Q$) measure:} is obtained using transition point tracking. Multiple transition points were tracked that were used to obtain the absolute deviation ($\delta$) between the original and predicted bifurcation point.

\begin{equation}
    Q = 1 - \overline{\delta}
\label{Eq_BifQuality}
\end{equation}

where $\overline{\delta}$ is the mean of deviations of different transition points namely; transition from period 1 to 2
($\delta_{12}$), from period 2 to 4 ($\delta_{24}$), beginning of chaos ($\delta_{Chaos-a}$) and the end of chaotic regime ($\delta_{Chaos-b}$).
$\overline{\delta}$ is also divided by the $\Delta F = |F_{max}-F_{min}|$ in order to normalize $Q\in[0,1]$. The period 4 trajectories do not exist in the original forced Duffing oscillator system for some values of $\omega$, see Fig.\ref{fig4}(e). Therefore, the transition $\delta_{24}$ was not considered for such values of $\omega$ while quantifying the quality $Q$ of the predicted bifurcation diagram.


\appendix

\section{Error in bifurcation points predicted from 1 and 2 samples over $f$.}

\begin{table*}
 \centering
\begin{tabular}{|c|c|c|c|}
\hline 
(Hyper)-parameters & Symbol & 1D & 2D \\ 
\hline
Number of reservoir nodes & $N$ & 100 & 100 \\ 
\hline
Reservoir network density & $\delta^{Res}$ & 0.5 & 0.8 \\ 
\hline 
Spectral radius of reservoir & $\rho$ & 0.5057 & 0.1057 \\ 
\hline 
Range of input connection weight & $\sigma$ & 0.2  & 0.2 \\ 
\hline 
Leaking parameter & $\alpha$ & 0.2 &  0.2\\ 
\hline 
Regularization parameter & $\beta$ & $5\times 10^{-8}$ & $5\times 10^{-8}$ \\ 
\hline 
Input ($f(t)$) length & $T$ & 6500 & 6500 \\
\hline
Transients removed before training & $T_{\text{trans}}$ & 400 & 400 \\
\hline
\end{tabular}
    \caption{List of parameters/hyper-parameters used for the numerical simulations.}
    \label{tab1}
\end{table*}

The detailed deviations of $\delta_{12}$, $\delta_{24}$, $\delta_{Chaos-a}$ and $\delta_{Chaos-b}$ marking the corresponding transition points in the 1D bifurcation diagram produced by auto-regressive-RCs trained over 1- and 2-training samples are shown in Fig.\ref{SupFig3_d}. These deviations are then used to calculate the overall Bifurcation Points Quality ($Q$) measure represented in Supplementary Fig.\ref{fig3}(e) given by Eq.\ref{Eq_BifQuality}. \\

\section{Parameters used in this study}

The forced Duffing oscillator was numerically integrated as an initial value problem using the SciPy \cite{SciPy} Python library, with a step-size of 0.1 and initial conditions $[q^{in}_1, q^{in}_2]=[0.05, 0.05]$.\\

For building the auto-regressive-RC models the list of parameters used in this study are shown in the Table \ref{tab1}.\\

The hyperparameters are optimized based on the Root Mean Square Error (RMSE) of the predicted signal. In this context, several hyperparameters are crucial for improving performance. One of the hyperparameter is number of reservoir nodes. Generally, increasing the number of reservoir nodes improves the quality of predictions. However, this improvement may saturate as the reservoir size continues to increase. Other hyperparameters include $\sigma$, $\alpha$, $\beta$ and $\rho$. The parameter $\alpha$ is the rate at which the reservoir state is updated. To maintain the echo state property, it is necessary that $\rho(W_{res}) < 1$. There are few references that explore hyperparameter tuning in detail \cite{lukovsevivcius2012practical}-\cite{jaeger2002tutorial}.


\nocite{*}


%

\end{document}